\documentclass{pnastwo}
\usepackage{t1enc}

\usepackage{graphicx}
\usepackage{color}
\usepackage{rotating}
\usepackage[dvips]{epsfig}
\usepackage{pnastwof}

\graphicspath{{}{images/}}

\def\Ref#1{{\bf \ref{#1}}}
\definecolor{ChannelDeposit}{rgb}{0.80000,  0.76078,  0.25882}
\definecolor{SandRidges}{rgb}{0.98824,  0.95686,  0.63529}
\definecolor{Swamp}{rgb}{0.65882,  0.81176,  0.21961}
\definecolor{Marsh}{rgb}{0.18824,  0.70588,  0.29020}
\definecolor{Marsh-mangroves}{rgb}{0.24706,  1.00000,  0.39216}
\definecolor{BeachRidges}{rgb}{0.98824,  0.79608,  0.63529}
\definecolor{EolianDunes}{rgb}{0.97647,  0.97647,  0.57647}
\definecolor{Floodplain}{rgb}{0.68627,  0.47059,  0.12549}
\definecolor{Uplands}{rgb}{0.71765,  0.65098,  0.54510}

\begin{document}

\conflictofinterest{Conflict of interest statement: No conflicts declared.}

\title{Modeling river delta formation} 
\author{Hansj\"org  Seybold
        \affil{1}{Computational Physics for Engineering Materials, IfB, 
                  ETH Zurich, 8093 Zurich, Switzerland},
 \thanks{To whom correspondence
should be addressed. E-mail: hseybold@ethz.ch}
        Jos\'e S.  Andrade Jr.
         \affil{2}{Departamento de F\'isica, 
                   Universidade Federal do Cear\'a, 60451-970 Fortaleza, 
                   Cear\'a, Brazil}\and
        Hans J. Herrmann\affil{1}{}\affil{2}{}  }

\maketitle

\begin{article}

\begin{abstract}
  A new model to simulate the time evolution of river delta formation
  process is presented.  It is based on the continuity equation for
  water and sediment flow and a phenomenological sedimentation/
  erosion law.  Different delta types are reproduced using different
  parameters and erosion rules.  The structures of the calculated
  patterns are analyzed in space and time and compared with real data
  patterns.  Furthermore our model is capable to simulate the rich
  dynamics related to the switching of the mouth of the river delta.
  The simulation results are then compared with geological records for
  the Mississippi river.
\end{abstract}

\keywords{River Deltas| Simulation | Fractals| Lattice Model}

\section{Introduction}\label{sec:Introduction}
The texture of the landscape and fluvial basins is the product of thousands of
years of tectonic movement coupled with erosion and weathering caused by water
flow and climatic processes.  To gain insight into the time evolution of the
topography, a model has to include the essential processes responsible for the
changes of the landscape.  In Geology the formation of river deltas and braided
river streams has been studied since a long time describing the schematic
processes for the formation of deltaic distributaries and inter-levee- basins
\cite{Coleman75, Allen81, Coleman88, Bridge93, Bristow93}.
Experimental investigation of erosion and deposition has a long tradition in
Geology \cite{Jaggar08}. Field studies have been carried out for the Mississippi
Delta \cite{Fisk47, Kolb58, Coleman64, Could70}, the Niger Delta
\cite{Allen64,Allen65, Allen70}, or for the Brahmaputra Delta \cite{Coleman69}.
Laboratory experiments have also been set up in the last decades for
quantitative measurements \cite{Czirok93, Ashmore82, Ashmore85, Wright05,
  Parker05, Parker03}.  For instance, in the eXperimental EarthScape (XES)
project the formation of river deltas is studied on laboratory scale and
different measurements have been carried out \cite{Kim06,Swendson05,Lague03}.

Nevertheless modeling has proved to be very difficult as the system is
highly complex and a large range of time scales is involved. To
simulate geological time scales the computation power is immense and
classical hydrodynamical models cannot be applied.  Typically these
models are based on a continuous \emph{ansatz} (e.g., shallow water
equations) which describes the interaction of the physical laws for
erosion, deposition and water flow \cite{Giacometti95, Willgoose91,
  Howard94, Kooi96, Densmore97, Beaumont00}. The resulting set of
partial differential equations are then solved with boundary and
initial conditions using classical finite element or finite volume
schemes.  Unfortunately none of these continuum models is able to
simulate realistic land-forms as the computational effort is much too
high to reproduce the necessary resolution over realistic time scales.
Therefore in the last years discrete models based on the idea of
cellular automata have been proposed \cite{Wolfram02, Murray94,
  Murray97, Davy00, Coulthard05, Coulthard06}.  These models consider
water input on some nodes of the lattice and look for the steepest
path in the landscape to distribute the flow. The sediment flow is
defined as a nonlinear function of the water flow and the erosion and
deposition are obtained by the difference of the sediment inflow and
outflow. This process is iterated to obtain the time evolution.  In
contrast to the former models, these models are fast and several
promising results have been obtained, but as they are only based on
flow, a well defined water level cannot be obtained with this
\emph{ansatz}.

Here we introduce a new kind of model where the water level and the landscape
are described on a lattice grid coupled by an erosion and sedimentation law.
The time evolution of the sediment and water flow is governed by conservation
equations.  The paper is organized as follows. After an overview on the
different types of deltas and their classification the model is introduced and
discussed in details.  The analysis of the model results and a comparison with
real landforms are provided.  According to different parameter combinations
different delta types can be reproduced and interesting phenomena in the time
evolution of a delta such as the switching of the delta lobe can be observed.
Finally the scaling structure of the delta pattern is analyzed and compared with
that obtained from satellite images.

\section{Classification}\label{sec:Classification}
The word ``delta'' comes from the Greek capital letter $\Delta$ and
can be defined as a coastal sedimentary deposit with both subaerial
and subaqueous parts. It is formed by riverborne sediment which is deposited
at the edge of a standing water. This is in most cases an ocean but
can also be a lake.  The morphology and sedimentary sequences of a
delta depend on the discharge regime and on the sediment load of the
river as well as on the relative magnitudes of tides, waves and
currents \cite{Coleman76}.  Also the sediment grain size and the water
depth at the depositional site are important for the shape of the
deltaic deposition patterns \cite{Coleman75, Coleman76,
  Bhattacharya92, Orton93}.  This complex interaction of different
processes and conditions results in a large variety of different
patterns according to the local situations.  Wright and Coleman
\cite{Coleman75, Coleman76, Wright73, Coleman73} described
depositional facies in deltaic sediments and concluded that they
result from a large variety of interacting dynamic processes (climate,
hydrologic characteristics, wave energy, tidal action, etc.)  which
modify and disperse the sediment transported by the river.  By
comparing sixteen deltas they found that the
Mississippi Delta is dominated by the sediment supply of the river
while the Senegal Delta or the S\~{a}o Francisco River Delta are mainly
dominated by the reworking wave activities. High tides and strong
tidal currents are the dominant forces at the Fly River Delta.

Galloway \cite{Galloway75} introduced a classification scheme where
three main types of deltas are distinguished according to the dominant
forces on the formation process: \emph{river-}, \emph{wave-} and
\emph{tide}-dominated deltas.  This simple classification scheme was later
extended \cite{Bhattacharya92, Orton93, Wright85} including also grain
size and other effects.

\begin{figure}
\includegraphics[width=7.5cm]{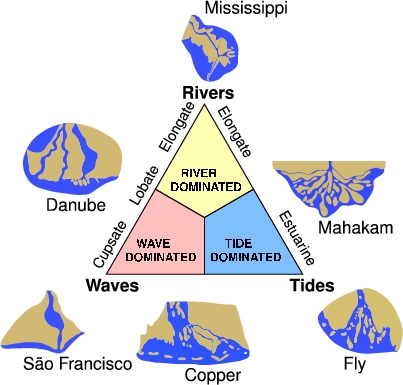}
\caption{The figure shows the classification scheme after Galloway
  \cite{Galloway75}, where wave- tide- and river-dominated deltas are
  distinguished in the extremes of the triangle.  By comparing 16 major river
  deltas Wright \& Coelman \cite{Wright73} concluded that in the extremes the
  Mississippi is the most river-dominated delta and the S\~{a}o Francisco the
  most wave-dominated one. The delta which is mainly dominated by the tides is
  that of the Fly river in Papua New Guinea.  }\label{fig:Classification}
\end{figure}

At the river-dominated end of the spectrum, deltas are indented and have more
distributaries with marshes, bays, or tidal flats in the interdistributary
regions.  They occur when the stream of the river and the resulting sediment
transport is strong and other effects such as reworking by waves or by tides are
minor \cite{Coleman76,Wright73}. These deltas tend to form big delta lobes into
the sea which may have little more than the distributary channel and its levee
is exposed above the sea level. Due to their similarity with a bird's foot, they
are often referred in the literature as ``bird foot delta'' like in the case of
the Mississippi River Delta \cite{Coleman76}.  When more of the flood plain
between the individual distributary channels is exposed above the sea level, the
delta displays lobate shape.  Wave-dominated delta shorelines are more regular,
assuming the form of gentle, arcuate protrusions, and beach ridges are more
common (e.g., like for the Nile Delta or Niger Delta \cite{Allen65, Oomkens74}).
Here the breaking waves cause an immediate mixing of fresh and salt water. Thus
the stream immediately loses its energy and deposits all its load along the
cost.  Tide-dominated deltas occur in locations of large tidal ranges or high
tidal current speeds. Such a delta often looks like a estuarine bay filled with
many stretched islands parallel to the main tidal flow and perpendicular to the
shore line (like e.g., the Brahmaputra Delta).  Using the classification of
Galloway \cite{Galloway75} the different delta types of deltas can be arranged
in a triangle where the extremes are put in the edges (see
Fig.~\ref{fig:Classification}).

\section{The Model}\label{sec:Model}
The model discretizes the landscape on an rectangular grid where the surface
elevation $H_i$ and the water level $V_i$ are assigned to the nodes. Both $H_i$
and $V_i$ are measured from a common base point, which is defined by the sea
level.  On the bonds between two neighboring nodes $i$ and $j$, a hydraulic
conductivity for the water flow from node $i$ to node $j$ is defined as
\begin{equation}
\label{eq:Sigma}
  \sigma_{ij}=c_\sigma\left\{
\begin{array}{cl}
\displaystyle{V_i+V_j\over2}-{H_i+H_j\over 2}&\mbox{if $>0$}\\
0 &\mbox{else}.
\end{array}
\right.
\end{equation}
As only surface water flow is considered, $\sigma_{ij}$ is set larger
than zero only if the water level of the source node is larger than
the topography, which means that water can only flow out of a node
where the water level is above the surface.  The relation between the
flux $I_{ij}$ along a bond and the water level is given by
\begin{equation}
\label{eq:Current}
  I_{ij}=\sigma_{ij}(V_i-V_j).
\end{equation}
Furthermore water is routed downhill using the continuity equation for
each node
\begin{equation}
  \label{eq:Continuity}
  \frac{V_i-V'_i}{\Delta t}=\sum_{N.N.} I_{ij},
\end{equation}
where the sum runs over all currents that enter or leave node $i$ and $V_i'$ is
the new water level.  The boundaries of the system are chosen as follows: On the
sea side the water level on the boundary is set equally to zero and water just
can flow out of the system domain.  On the land the water is retained in the
system by high walls or choosing the computational domain for the terrain such
that the flow never reaches the boundary.  Water is injected into the system by
defining an input current $I_0$ at the entrance node.

The landscape is initialized with a given ground water table.  Runoff
is produced when the water level exceeds the surface.  The sediment
transport is coupled to the water flow by the rule, that all sediment
that enters a node has to be distributed to the outflows according to
the strength of the corresponding water outflow. Thus the sediment
outflow currents for node $i$ are determined via
\begin{equation}
\label{eq:Sediment_flow}
  J^{out}_{ij}={\sum_k J^{in}_{ik}\over \sum_k |I^{out}_{ik}|}I^{out}_{ij},
\end{equation}
where the upper sum runs over all inflowing sediment and the lower one
over the water outflow currents.  A sediment input current $s_0$ is
defined in the initial bond.

The sedimentation and erosion process is modeled by a phenomenological
relation which is based on the flow strength $I_{ij}$ and the local
pressure gradient imposed by the difference in the water levels in the
two nodes $V_i$ and $V_j$. The sedimentation/erosion rate $dS_{ij}$
is defined through
\begin{equation}
  \label{eq:Erosion}
  dS_{ij}=c_1(I^\star-|I_{ij}|)+c_2(V^\star-|V_i-V_j|),
\end{equation}
where the parameters $I^\star$ and $V^\star$ are erosion thresholds and the
coefficients $c_1$ resp. $c_2$ determine the strength of the corresponding
process.  The first term $c_1(I^\star-|I_{ij}|)$ describes the dependency on the
flow strength $I_{ij}$ \cite{Wipple99} and is widely used in geomorphology,
while the second term $c_2(V^\star-|V_i-V_j|)$ relates sedimentation and erosion
to the flow velocity, which in the model can be described by
$I_{ij}/\sigma_{ij}\sim|V_i-V_j|$.  The two terms of \Ref{eq:Erosion} are not
linearly dependant on each other as one may think first by looking at Eq.
\Ref{eq:Current}.  In fact due to Eq.\Ref{eq:Sigma} there is a nonlinear
relation between $V$ and $I$ which leads to different thresholds in the pressure
gradient and the current.

The sedimentation
rate $dS_{ij}$ is limited by the sediment supply through $J_{ij}$,
thus in the case $dS_{ij}>J_{ij}$ the whole sediment is deposited on
the ground and $J_{ij}$ is set to zero. In the other cases $J_{ij}$ is
reduced by the sedimentation rate or increased if we have erosion. The
erosion process is also supply limited which means that the erosion
rate is not allowed to exceed a certain threshold $T$; so $\mbox{if }
dS_{ij}<T, \mbox{ then } dS'_{ij}=T$. Note that in the case of erosion
$dS_{ij}$ is negative.  Due to erosion or deposition, the landscape is
modified according to
\begin{eqnarray}
  \label{eq:Landscape_Modification1}
   H'_i&=&H_i+\frac{\Delta t}{2}dS_{ij}\\
    \label{eq:Landscape_Modification2}
   H'_j&=&H_j+\frac{\Delta t}{2}dS_{ij},
\end{eqnarray}
where the sediment deposits equally on both ends of the bond.  The new
topography is marked with $H'_i$. The same formulae (Eqs.
\Ref{eq:Landscape_Modification1}, \Ref{eq:Landscape_Modification2}) also hold in
the case of erosion when $dS_{ij}$ is negative. 

Iterating Eqns. \Ref{eq:Sigma} to \Ref{eq:Landscape_Modification2} determines
the time evolution of the system. Finally in a real system subaquateous water
currents lead to a smoothening of the bottom which is modeled by the following
expression
\begin{equation}
  \label{eq:smooth}
  H'_i= (1-\epsilon)H_i+\frac{\epsilon}{4}\sum_{N.N.}H_j,
\end{equation}
where $\epsilon$ is a smoothening constant determining the strength of
the smoothening process. The sum runs over all nearest neighbors of
node $i$.

\section{Simulation}\label{sec:Simulation}

\begin{figure}
  \includegraphics[angle=0,width=7.5cm]{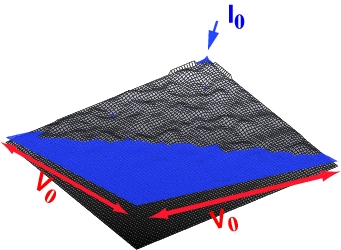}
  \caption{The figure shows a sketch of the initial condition for a simulation.
    A water current $I_0$ is injected at the upper node and the water levels on
    the sea boundaries are kept constant ($V_0=0$). The landscape is initialized
    as an inclined plane with a disordered topography on the top. The water
    surface (blue) is parallel to the horizontal plane.
  }\label{fig:Initialization}
\end{figure}

\begin{figure}
\includegraphics[width=18cm]{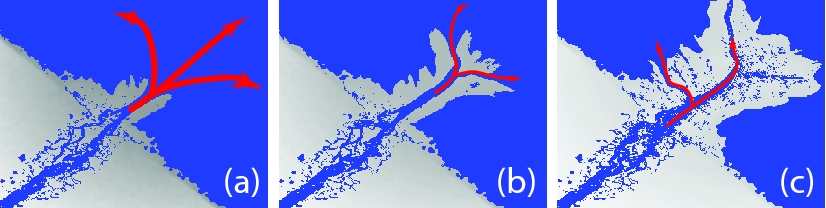}
\caption{Time evolution of a birdfoot delta (from left to right).  Figure (a)
  shows the delta after 1.2 million time steps, where the main channel worked
  into the sea depositing sediment mainly on its levee sides.  After 2.5 million
  time steps the main channel has split into two distributaries (b), where the
  smaller one becomes inactive after 5 million steps and a new channel breaks
  through the sidewalls (c). The main directions of the sediment flow are marked
  with the red arrows.  The simulation was run on a 279x279 lattice and the
  parameters for the water flow were $I_0=1.7\times 10^{-4}$ and $c_\sigma=8.5$.
  For the sedimentation and erosion the constants were set to $c_1=0.1$ and
  $c_2=0$ with a sediment input current of $s_0=0.00025$.  The erosion threshold
  $I^\star$ was set to $I^\star=4\times 10^{-6}$ and the maximal erosion rate
  was set to $|T|=5\times 10^{-7}$.  Smoothening was applied every 2000 time
  steps with a smoothening factor of $\epsilon=1\times 10^{-4}$.  The initial
  depth of the water table at the bottom of the valley was set to
  $\delta=0.0025$.  }
\label{fig:Birdfoot_simu1}
\end{figure}

The simulation is initialized with a valley on a rectangular $N\times N$ lattice
with equal spacing grid as shown in Fig.~\ref{fig:Initialization}. The valley
runs downhill with slope $S$ along the diagonal of the lattice and the
hillslopes of the valley increase from the bottom of the valley sidewards
according to a power law with exponent $\alpha$.  In the simulation shown in
Fig.~\ref{fig:Birdfoot_simu1} the value of $\alpha$ was chosen to be 2.0.  Under
the sea the landscape is flat with a constant slope downhill.  Furthermore we
assume the initial landscape to have a disordered topography by assigning
uniformly distributed random numbers to $H_i$.  This variable is then smoothed
out according to Eq.~\Ref{eq:smooth}.  The water level $V_i$ of the system is
initialized with a given ground water table.  In reality the distance of the
ground water to the surface is minimal on the bottom of the valley and increases
uphill.  This is obtained in the simulation by choosing the water level $V_i$ in
an incline plane $\delta$ below the bottom of the valley.  The slope of the
plane is the same as the slope of the valley $S$.
This also keeps the river close to the bottom of the valley.  As we are only
interested in studying the pattern formation at the mouth of the river, the
braiding conditions of the upper river only determine the width of the delta
front.  On the seaside when $H_i\leq 0$ the water level is a constant and set to
zero.  A sketch of the initial landscape is shown in
Fig.~\ref{fig:Initialization}.

An initial channel network is created by running the algorithm without
sedimentation and erosion until the water flow reaches a steady state.  Then the
sedimentation and erosion procedure is switched on and the pattern formation at
the mouth of the river is studied.

According to the dominance of the different processes, completely different
coastline shapes can be observed. The smoothening procedure Eq.~\Ref{eq:smooth}
leads to the formation of an estuary by reworking the coastline at the river
mouth, while the stream dominant erosion term $c_1(I^\star-|I_{ij}|)$ in
Eq.~\Ref{eq:Erosion} favors the formation of river-dominated birdfoot shaped
delta.  In contrast to this, the second term $c_2(V^\star-|V_i-V_j|)$ in
Eq.~\Ref{eq:Erosion}, which depends on the pressure gradient represented by the
height difference of the water levels in the nodes $i$ and $j$, produces more
classical shaped deltas with several islands and channels. These patterns are
similar to the distributary structure of the Lena or the Mahakam river delta,
which are more sea or wave-dominated.  This difference can be explained by the
fact that the first term sediments along the main current stream, whereas the
second term distributes the sediment more equally to the neighboring nodes.

\begin{figure}
\includegraphics[angle=0,width=7.5cm]{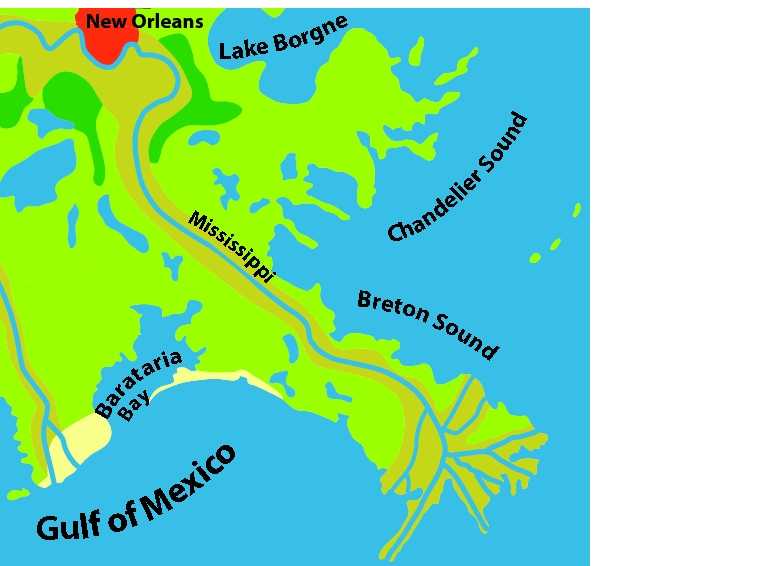}
\caption{For comparison with the simulation results of
  Fig.~\ref{fig:Birdfoot_simu1} the figure shows part of a map of the mouth of
  the Mississippi river, where the birdfoot shaped delta can be seen clearly.
  The colors indicate channel deposits 
  \colorbox{ChannelDeposit}{\textcolor{ChannelDeposit}{x}}, 
  sand ridges \colorbox{SandRidges}{\textcolor{SandRidges}{x}}, 
  swamps \colorbox{Swamp}{\textcolor{Swamp}{x}} and 
  marshes \colorbox{Marsh}{\textcolor{Marsh}{x}}.
  The figure was generated after \cite{Coleman76}.}\label{fig:Mississippi}
 
\end{figure}

Figures \ref{fig:Birdfoot_simu1}(a-c) show some snapshots of the time evolution
of the simulation of a birdfoot delta ($c_2=0$). A map of the Mississippi River
is given in Fig.~\ref{fig:Mississippi} for comparison.  In both cases one can
see how the main channel penetrated into the ocean depositing sediment mainly on
its levee sides.  When the strength of the main channel decreases, side channels
start to appear breaking through the sidebars as can be seen in the snapshots of
Figs.~\ref{fig:Birdfoot_simu1}(b) and ~\ref{fig:Birdfoot_simu1}(c). At the
beginning of the delta formation process the sediment transport is equally
distributed among the different channels and leads to a broader growth of the
delta front along the coast.  With time, the side channels are gradually
abandoned and the sediment is primarily routed through the main channel, thus
this dominant channel is growing faster than the others forming the typical
birdfoot shaped deposits.

\begin{figure}
\includegraphics[angle=0,height=5cm]{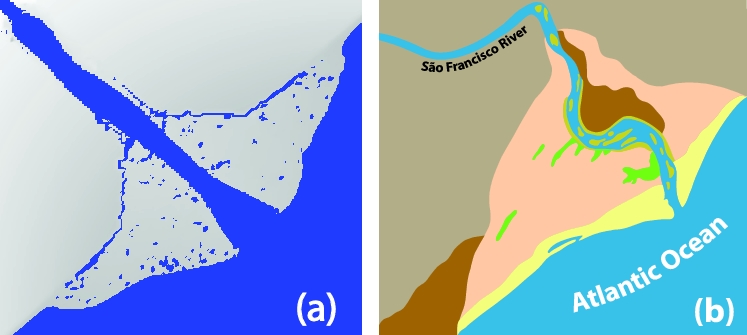}
\caption{In (a) the simulation of a wave-dominated delta is shown.  While the
  waves are reworking the coast at the mouth of the river to form an estuary,
  the river deposits sediment and forms large beaches.  As the simulation does
  not include evaporation, the ponds and inactive channels in the deposition
  zone do not disappear as in the map of the real river shown in (b). The
  parameters in the simulation were $N=179$, $I_0=1.7\times 10^{-4}$,
  $s_0=0.0015$, $c_\sigma=8.5$, $c_1=0$, $c_2=0.1$ and $I^\star=1.3\times
  10^{-4}$. Smoothening was applied every 200 time steps with a smoothening
  constant $\epsilon=0.01$. In (b) we show for comparison a map of the S\~{a}o
  Francisco river delta in southern Brazil which is the most wave-dominated
  delta according to the classification of \cite{Galloway75}.
  The colors in the map (Fig.~(b))  indicate channel deposits
   \colorbox{ChannelDeposit}{\textcolor{ChannelDeposit}{x}}, 
   beach ridges \colorbox{BeachRidges}{\textcolor{BeachRidges}{x}}, 
   eolian dunes \colorbox{EolianDunes}{\textcolor{EolianDunes}{x}},
   marsh-mangroves \colorbox{Marsh-mangroves}{\textcolor{Marsh-mangroves}{x}},
   the floodplain \colorbox{Floodplain}{\textcolor{Floodplain}{x}} and the uplands
\colorbox{Uplands}{\textcolor{Uplands}{x}}.
  The figure was  generated after \cite{Coleman76}. }\label{fig:Wave_simu}
\end{figure}

Figure \ref{fig:Wave_simu}(a) shows another type of delta where the smoothening
of the waves reworks the deposits at the river mouth and distributes it along
the coast. Here the river could built up only a slight protrusion in the
immediate vicinity of the river mouth. The same happens in areas where the wave
currents are dominant and lead to the formation of wave-dominated deltas like
the S\~{a}o Francisco River in Brazil or the Nile Delta. A map of the S\~{a}o
Francisco River Delta is also given in Fig.~\ref{fig:Wave_simu}(b) for
comparison.  Here the coast line has been straightened by the wave activities
and consists almost completely of beach ridges which have the typical triangular
shape inland.  This flattened deposit can also be found in our simulation
results.  As there is no evaporation included in the simulation, small ponds and
abandoned channels remain in the sedimented zone instead of disappearing with
time.

Finally, if the term $c_2(V^\star-|V_i-V_j|)$ dominates the
sedimentation/erosion process a half moon shaped delta with many small islands
and channels appear.  This delta type shows more activity in the channel network
than the others.  The channels split and come together, and when the main
channel blocks its way due to sedimentation, the whole delta lobe switches to
another place.  This phenomenon is called delta switching.  During the
simulation, the switching of the delta occurred several times.

\begin{figure}
\includegraphics[height=4.2cm]{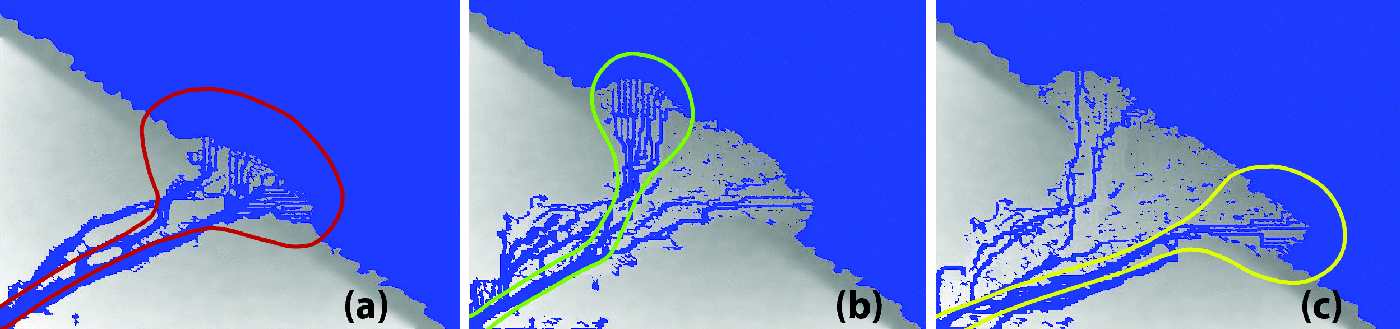}
\caption{The figures show the switching of the delta lobe during the simulation.
  Comparing Figs.~(a) and (b) a type I switching can be identified where the
  main part of the delta lobe is abandoned close to the mouth of the river just
  before the river splits into several distributaries and forms a new lobe
  beside.  Another type of delta switching is shown comparing Figs.~(b) and (c)
  with two snapshots from the simulation. Here the channel switches far upstream
  and takes a new course to the coast forming another delta lobe far away.  This
  is referred as a switching of type II.  The parameters for the simulation
  where $I_0=1.7\times 10^{-4}$, $s_0=5\times 10^{-5}$, $c_2=0.0005$, $c_1=0$
  and $I^\star=3.3\times 10^{-4}$. The simulation was run on a $179\times179$
  lattice with smoothening every 2000 time steps and a smoothening constant of
  $\epsilon=0.0001$. }\label{fig:SwitchingTypeI}
\end{figure}

The best studied delta in the world is that of the Mississippi river where the
switching of the delta lobes was studied in detail \cite{Kolb58}. The switching
of the Mississippi Delta during the last 4000 years is well documented
\cite{Fisk47,Kolb58, Fisk52}.
The rich dynamics due to the switching phenomenon that is observed in the
Mississippi can be also identified in our simulations.  In the literature
\cite{Coleman76} three types of switching mechanisms are distinguished. The
first type referred as switching type I, consists of a lobe switching in which
the delta propagates in a series of distributary channels. After a certain time,
the stream abandons the entire system close to the head of the delta and forms a
new lobe in an adjacent region.  Very often this lobe occupies an indentation in
the coastline between previous existing lobes so that with time the sediment
layers overlap each other.  One can find this type of delta switching in areas
where the offshore slope is extremely low and the tidal and wave forces are too
small for reworking the lobe \cite{Coleman76, Wright73, Galloway75}.  In many
cases the delta lobes merge with each other forming major sheet-type sand banks.
This phenomena can be nicely observed when comparing the two images of the
simulation in Figs.~\ref{fig:SwitchingTypeI}(a) and \ref{fig:SwitchingTypeI}(b).
In the Mississippi Delta this also happened several times in the past and the
different lobes have today different names.  A type I shift of the Mississippi
delta occurred for example, 4600 years B.C. between the Sal{\'e}-Cypremort and
the Cocodrie (4600-3500) Lobe or when the St.  Bernard Lobe switched to which
today is called Lafourche at about 1000 BC.

At about 3500 BC the Mississippi river switched far upstream from the Cocodrie
to the Teche stream tailing a completely new course for the river itself and its
delta.  This type of switching is referred as type II switching \cite{Coleman76}
and can also be found in the simulation. When comparing
Figs.~\ref{fig:SwitchingTypeI}(b) and \ref{fig:SwitchingTypeI}(c) one can see a
major shift of the channel far upstream in the deltaic plain so the river takes
a completely different course and forms a new delta beside.

The type III of delta switching is referred in the literature as alternate
channel extension \cite{Coleman76}. In this case not the complete channel but
the dominance of sediment flux in one or more distributaries is changing with
time.  This can be described as follows: two or more major channels split into
several distributaries nearly at the same point at the head of the delta.
Commonly one of the distributaries is dominant, so it will carry most of the
sediment and water discharge at any time.  As a result, this active channel will
rapidly propagate seaward, while the other channel will shrivel with time.  At
some point, the slope of the main active channel will decrease and the discharge
will seek one of the shorter distributaries. With the increased sediment flux
downstream, the new channel will rapidly propagate into the sea.  This switching
process will repeat several times forming a deltaic plain characterized by a
series of multiple beach ridges.  This type of switching can be best observed in
the simulation of the birdfoot delta in Fig.~\ref{fig:Birdfoot_simu1}(a-c),
where the main path of the sediment flow is marked by the red arrows. One can
see how side channels emerge and are abandoned after a certain time.
Nevertheless, a major switching of the main channel could not be observed in the
simulations.  The average time between two lobe switchings was found to be
around 1000 years for the Mississippi river \cite{ Kolb58, Coleman76, Fisk52}.

At this point, we show that the river delta patterns generated from our
simulations display geometric features that are statistically similar to real
river delta structures.  More precisely, we analyze the self similar behavior of
the real and simulated river deltas using the box counting algorithm
\cite{Feder89}.  The box counting dimension is a quite common measure in
geomorphological pattern analysis and has been used by many authors to
characterize river basin patterns and coastlines 
\cite{Maritan96, Rodriguez-Iturbe98, Sapoval04}.

\begin{figure}
\includegraphics[width=17cm]{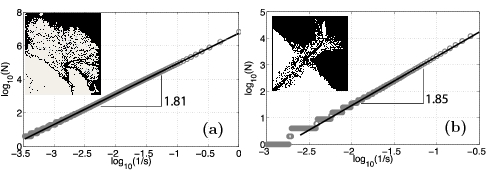}
\caption{ Figure (a) shows the scaling behavior of the Lena river delta.  On the
  y-axis we show the logarithm of the number of boxes $N(s)$ of size $s$ which
  are necessary to cover the subaerial surface is plotted versus the logarithm
  of the inverse box size.  The straight line is a power law fit $N\sim s^{-D}$
  with exponent $D=1.81$.  In the inset a satellite picture of the Lena delta is
  shown.  In (b) one can see the scaling behavior of the birdfoot delta from the
  simulation (c.f. Fig.~\ref{fig:Birdfoot_simu1}) where the slope was calculated
  to be 1.85. }\label{fig:Fractal}
\end{figure}

For the real satellite picture as well as for
the simulated river delta, we show in Figs.~\ref{fig:Fractal} that the variation
with the cell size $s$ of the number of cells $N$ covering the land follows
typical power laws over more than 3 decades
\begin{equation}
  \label{eq:fractal}
  N\sim s^{-D},
\end{equation}
where the exponent $D$ is the fractal dimension.  Moreover, the least square fit
of this scaling function to the data gives exponents which are strikingly close
to each other, namely $D=1.81\pm 0.01$ for the real Lena river delta and
$D=1.85\pm 0.1$ for the simulation.

\section{Conclusion}\label{sec:Conclusion}
A new model for simulating the formation process of river deltas has been
presented. It is based on simple conservation laws for water and sediment on a
lattice grid, coupled by a phenomenological sedimentation/erosion law.  Several
interesting features of river deltas like the different delta switching
processes could be found with the model and compared with real landforms.

Different delta shapes in the classification scheme of Galloway
\cite{Galloway75} could be reproduced by varying the model parameters and
initial conditions. The pattern structure of the simulation has been analyzed
and good agreement with real deltas have been found.  Furthermore the delta
shifting phenomena could be observed in the simulation and different types of
delta shifting could be distinguished.

\begin{acknowledgments}
  This work was funded by the Swiss National Fond and the Max Planck Prize as
  well as by the Landesstiftung BW, CNPq, CAPES and FUNCAP.
\end{acknowledgments}
\bibliographystyle{pnas}
\bibliography{RiverNetworks} 

\begin{thebibliography}{10}

\bibitem{Coleman75}
Coleman, J \& Wright, L.
\newblock (1975) {\em Modern river deltas: variability of processes and sand
  bodies} ed.{} Broussard, M.
\newblock (Houston Geological Society), pp. 99--149.

\bibitem{Allen81}
Allen, G, Laurier, D,  \& Thouvenin, J.~P.
\newblock (1981) {\em A.A.G.P. Bull.} {\bf 65}, 889--889.

\bibitem{Coleman88}
Coleman, J.
\newblock (1988) {\em Geol. Soc. Am. Bull.} {\bf 100}, 999--1015.

\bibitem{Bridge93}
Bridge, J.
\newblock (1993) in {\em Braided Rivers}, ed.{} Bristow, C.
\newblock (Geol. Soc., London), pp. 13--71.

\bibitem{Bristow93}
Bristow, C \& Best, J.
\newblock (1993) in {\em Braided Rivers}, ed.{} Bristow, C.
\newblock (Geol. Soc., London), pp. 1--11.

\bibitem{Jaggar08}
Jaggar, T.~A.
\newblock (1908) {\em Bull. Mus. Compar. Zool.} {\bf 49}, 285--305.

\bibitem{Fisk47}
Fisk, H.
\newblock (1947) Fine-grained alluvial deposits and their effects on
  mississippi river activities, (U.S. Army Corps of Engineering, Mississippi
  River Commission), Technical report.
\newblock http://lmvmapping.erdc.usace.army.mil/.

\bibitem{Kolb58}
Kolb, C.~R \& Lopik, J.~v.
\newblock (1958) Geology of the mississippi deltaic plain, southeastern
  louisiana, (U.S. Army Corps of Engineering, Waterways Experiment Station),
  Technical report.

\bibitem{Coleman64}
Coleman, J \& Gagliano, S.
\newblock (1964) {\em Gulf Clast Assn. Geol. Soc. Trans.} {\bf 14}, 67--80.

\bibitem{Could70}
Could, H.
\newblock (1970) {\em The Mississippi Delta complex} ed.{} Morgan, J.
\newblock (Soc. of Economic Paleontologists and Mineralologists Special
  Publication), Vol.{}~15, pp. 3--30.

\bibitem{Allen64}
Allen, G.
\newblock (1964) {\em Mar. Geol.} {\bf 1}, 289--332.

\bibitem{Allen65}
Allen, G.
\newblock (1965) {\em Geol. en Mijnb} {\bf 44}, 1--21.

\bibitem{Allen70}
Allen, G.
\newblock (1970) {\em Sediments of the modern Niger Delta} ed.{} Morgan, J.
\newblock pp. 138--151.

\bibitem{Coleman69}
Coleman, J.
\newblock (1969) {\em Sediment. Geol.} {\bf 5}, 39--57.

\bibitem{Czirok93}
Czir\'ok, A \& Somfai, E.
\newblock (1993) {\em Phys. Rev. Lett.} {\bf 71}, 2154--2157.

\bibitem{Ashmore82}
Ashmore, P.
\newblock (1982) {\em Earth Surf. Proc. Land.} {\bf 7}, 201--225.

\bibitem{Ashmore85}
Ashmore, P.
\newblock (1985) Ph.D. thesis (University of Alberta, Alberta).

\bibitem{Wright05}
Wright, S \& Parker, G.
\newblock (2005) {\em J. Hydr. Res.} {\bf 43}, 612--630.

\bibitem{Parker05}
Parker, G.
\newblock (2005) Sediment transport morphodynamics, with applications to
  fluvial and subaqueous fans and fan-deltas (edited, copyrighted e-book).

\bibitem{Parker03}
Parker, G, Toro-Escobar, M, Ramey, M,  \& Beck, S.
\newblock (2003) {\em J. Hydraul. Eng.} {\bf 129}, 885--895.

\bibitem{Kim06}
Kim, W, Paola, C, Voller, V,  \& Swenson, J.
\newblock (2006) {\em J. Sediment. Res.} {\bf 76}, 270--283.

\bibitem{Swendson05}
Swenson, C, Paola, C, Pratson, L, Voller, V.~R,  \& Murray, A.~B.
\newblock (2005) {\em J. Geophys. Res.} {\bf 110}, 1--16.

\bibitem{Lague03}
Lague, D, Crave, A,  \& Davy, P.
\newblock (2003) {\em J. Geophys. Res.} {\bf 108}.
\newblock doi: 10.1029/2002JB001785.

\bibitem{Giacometti95}
Giacometti, A, Maritan, A,  \& Banavar, J.
\newblock (1995) {\em Phys. Rev. Lett.} pp. 577--580.

\bibitem{Willgoose91}
Willgoose, G, Bras, R,  \& Rodriguez-Iturbe, I.
\newblock (1991) {\em Water Resour. Res.} {\bf 27}, 1671--1684.

\bibitem{Howard94}
Howard, A.~D.
\newblock (1994) {\em Water Resour. Res.} {\bf 30}, 2261--2285.

\bibitem{Kooi96}
Kooi, H \& Beaumont, C.
\newblock (1996) {\em J. Geophys. Res.} {\bf 101}, 3361--3386.

\bibitem{Densmore97}
Densmore, A, Anderson, R, McAdoo, B,  \& Ellis, M.
\newblock (1997) {\em Science} {\bf 275}, 369--372.

\bibitem{Beaumont00}
Beaumont, C, Kooi, H,  \& Willett, C.
\newblock (2000) {\em Coupled tectonic-surface process models with applications
  to rifted margins and collisional orogens} ed.{} Summerfield, M.
\newblock (Wiley), pp. 29--55.

\bibitem{Wolfram02}
Wolfram, S.
\newblock (2002) {\em A new kind of science}.
\newblock (Wolfram Media Inc., Champaign, Illinois, USA).

\bibitem{Murray94}
Murray, A \& Paola, C.
\newblock (1994) {\em Nature} {\bf 371}, 54--57.

\bibitem{Murray97}
Murray, A \& Paola, C.
\newblock (1997) {\em Earth. Surf. Proc. Land.} {\bf 22}, 1001--1025.

\bibitem{Davy00}
Darvy, P \& Carve, A.
\newblock (2000) {\em Phys. Chem. Earth} {\bf 25}, 533--541.

\bibitem{Coulthard05}
Coulthard, T.~J.
\newblock (2005) {\em Water Resour. Res.} {\bf 41}.
\newblock 04003, doi:10.1029/2004WR003201.

\bibitem{Coulthard06}
Coulthard, T.~J \& van~de Viel, M.~J.
\newblock (2006) {\em Earth. Surf. Proc. Land.} {\bf 31}, 123--132.

\bibitem{Coleman76}
Coleman, J.
\newblock (1975) {\em Deltas: Processes of Deposition and Models for
  Exploaration}.
\newblock (Continuing Education Publication Company, Inc., Campain, IL).

\bibitem{Bhattacharya92}
Bhattacharya, J.~P \& Walker, R.~G.
\newblock (1992) in {\em Facies Models, Response to Sea-Level Change}, eds.{}
  Walker, R.~G \& James, N.
\newblock (Geological Association of Canada, St. Johns), pp. 157--177.

\bibitem{Orton93}
Orton, G.~J \& Reading, H.~G.
\newblock (1993) {\em Sedimentology} {\bf 40}, 475--512.

\bibitem{Wright73}
Wright, L.~D \& Coleman, J.~M.
\newblock (1973) {\em A.A.P.G. Bull.} {\bf 57}, 370--398.

\bibitem{Coleman73}
Coleman, J \& Wright, L.
\newblock (1973) {\em Trans. Gulf Coast Assoc. Geol. Soc.} {\bf 23}, 33--36.

\bibitem{Galloway75}
Galloway, W.
\newblock (1975) {\em Process framework for describing the morphologic and
  stratigraphic evolution of deltaic depositional systems} ed.{} Broussard, M.
\newblock (Houston Geological Society), pp. 87--98.

\bibitem{Wright85}
Wright, L.
\newblock (1985) in {\em River deltas. Coastal Sedimentary Environments}, ed.{}
  Davis, R. A.~J.
\newblock (Springer-Verlag, New York), pp. 1--76.

\bibitem{Oomkens74}
Oomkens, E.
\newblock (1974) {\em Sedimentology} {\bf 21}, 145--222.

\bibitem{Wipple99}
Wipple, K \& Tucker, G.
\newblock (1999) {\em J. Geophys. Res.} {\bf 104}, 17661--17667.

\bibitem{Fisk52}
Fisk, H.
\newblock (1952) Geological investigation of the atchafalaya basin and the
  problem of mississippi river diversion, (U.S. Army Corps of Engineering,
  Mississippi River Commission), Technical report.
\newblock http://lmvmapping.erdc.usace.army.mil/.

\bibitem{Feder89}
Feder, J.
\newblock (1989) {\em Fractals}.
\newblock (Plenum Press, New York), 4 edition.

\bibitem{Maritan96}
A.~Maritan, A, Colaiori, F, Flammini, A, Cieplak, M,  \& Banavar, J.
\newblock (1996) {\em Science} {\bf 272}.

\bibitem{Rodriguez-Iturbe98}
Rodriguez-Iturbe, I \& Rinaldo, A.
\newblock (1997) {\em Fractal River Basins: Chance and Self-Organization}.
\newblock (Cambridge University Press, New Youk).

\bibitem{Sapoval04}
Sapoval, B, Baldassarri, A,  \& Gabrielli, A.
\newblock (2004) {\em Phys. Rev. Lett.}

\end{thebibliography}

\end{article}

\newpage

\end{document}